\documentclass[twocolumn,letter]{jpsj2}
\usepackage{times}

\def\runauthor{%
H. \textsc{Yoshioka}, M. \textsc{Tsuchiizu}, 
 and H. \textsc{Seo}}

\title{%
Charge-Ordered State versus Dimer-Mott Insulator at Finite Temperatures
}

\author{%
Hideo \textsc{Yoshioka}%
  $^1$\thanks{E-mail address: h-yoshi@cc.nara-wu.ac.jp},
Masahisa \textsc{Tsuchiizu}%
  $^2$\thanks{E-mail address: tsuchiiz@slab.phys.nagoya-u.ac.jp}, and
Hitoshi \textsc{Seo}%
  $^3$\thanks{E-mail address: seo0@spring8.or.jp} 
}

\inst{%
$^1$Department of Physics, Nara Women's University, Nara 630-8506 \\
$^2$Department of Physics, Nagoya University, Nagoya 464-8602 \\
$^3$Synchrotron Radiation Research Center, Japan Atomic Energy Agency,
SPring-8, Hyogo 679-5148
}

\recdate{\today}

\abst{%
We theoretically investigate the competition 
 between charge-ordered state and
 Mott insulating state at finite temperatures
 in quarter-filled quasi-one-dimensional electron systems,  
 by studying 
 dimerized extended Hubbard chains 
 with interchain Coulomb interactions.  
In order to take into account one-dimensional fluctuations properly,
 we apply the bosonization method to an effective model
 obtained by the interchain mean-field approximation.
The results show 
that lattice dimerization, especially in the critical region, 
and frustration in the interchain Coulomb interactions 
 reduce the charge-ordering phase transition temperature 
 and enlarge the dimer-Mott insulating phase. 
We also derive a general formula of the Knight shift
  in the charge-ordered phase
 and its implication to experiments is discussed.
}

\kword{%
charge-ordering, 
quarter filling, quasi one-dimension,
extended Hubbard model, lattice dimerization, 
Mott insulator, molecular conductors
}

\begin{document}
\maketitle

Since the experimental observations of the charge-ordered (CO) state 
 in quasi-one-dimensional (1D) organic 2:1 compounds\cite{Review}, e.g.,
 (DI-DCNQI)$_2$Ag\cite{Hiraki1998PRL}
 and (TMTTF)$_2X$\cite{Chow2000PRL,Monceau2001PRL}
($X$: a monovalent counter anion), 
 in which the charge distribution becomes 
 disproportionated in a regular way 
 due to the electron-electron Coulomb repulsion, 
 these systems have been extensively studied, 
   experimentally\cite{TakahashiReview}
  as well as theoretically\cite{SeoReview}. 
In (TMTTF)$_2X$, the CO state is easily destroyed by
 external pressure\cite{Zamborszky2002PRB,Yu2004PRB} 
 but nevertheless the system remains insulating, 
 which is due to a competing state, 
 i.e., the so-called dimer-Mott insulating state 
 where the charge localizes on each lattice-dimerized bond. 
On the other hand, in (DI-DCNQI)$_2$Ag, 
 a recent X-ray structural analysis\cite{Kakiuchi2007PRL}
  revealed the existence of a three-dimensional mixtured pattern of CO and bond dimerization 
 in the ``CO phase''. 
In this paper, we theoretically investigate such interplay between 
 the CO state and the lattice dimerization based on a quasi-one-dimensional model, 
 especially their properties at \textit{finite-temperatures}.

Previous theoretical studies on the CO phenomena in these quasi-1D materials 
 have mainly been devoted to ground-state properties of 
  the 1D quarter-filled extended Hubbard model (EHM) with 
  the on-site and nearest-neighbor Coulomb repulsions $U$ and $V$ 
as a minimal model.\cite{SeoReview} 
In this model,  CO insulator (COI) is stabilized at $T=0$ 
  in the large $U$ and $V$ region, 
 while the system is in the metallic state described 
  by the Tomonaga-Luttinger liquid (TLL)  otherwise 
 (for positive $U$ and $V$)~\cite{Mila93EPL,Yoshioka00JPSJ,Ejima05EPL}. 
When the lattice dimerization is included by alternation in the transfer
  integrals, 
 the TLL phase is transformed to the dimer-Mott insulating phase, 
 and this competes with the CO state.~\cite{Seo97JPSJ,Ejima2006PRB,noteCODM}
There, the fluctuation effects in the competing region 
 have been focused based on the bosonization theory~\cite{Tsuchiizu01JPSJ} and 
 critical fluctuation appears on the boundary 
    between the two competing states 
    \cite{Fabrizio00NP,Tsuchiizu02JPCS,Otsuka04PRB}
  where its universality class is the two-dimensional Ising-type.

However,
  these results cannot be directly applied to the analysis of
  the actual materials,
  since purely 1D electronic models do not show any phase transition 
  at finite temperatures.  
In our previous work~\cite{Yoshioka06JPSJ}, 
 in addition to the 1D quarter-filled EHM, 
 we have taken into account the interchain Coulomb interaction $V_\perp$
 and investigated the CO phase transition 
  whose transition temperature $T_{\rm CO}$ 
 becomes always finite due to the dimensionality effect. 
We found that the interchain interaction can transform  
 the system from TLL metallic at $V_\perp=0$ to CO insulating by infinitesimal $V_\perp\neq0$
 in the critical region 
 and thus the interchain interaction greatly enhances the COI phase in the $U$-$V$ plane.
 
%------------------- Fig.1 ----------------------------------------------------
\begin{figure}[t]
\begin{center}\leavevmode
\includegraphics[width=6.5cm]{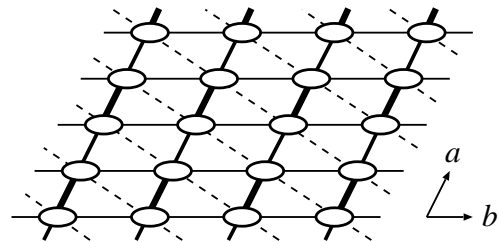}
\end{center}
\caption{
Schematic illustration of the model investigated in the present study.
Each ellipse denotes a molecular site and the $a$ ($b$) axis shows the
 chain (interchain) direction.
The thick (thin) lines along the $a$-axis shows the
  lattice dimerization of the stronger (weaker) bonds.  
The solid and dashed lines connecting the chains represent 
 the interchain interactions, $V_\perp$  and $V_\perp'$, respectively.         
}
\label{fig1}
\end{figure}
%-----------------------------------------------------------------------

In the present study, we consider a system 
shown schematically in Fig. \ref{fig1}, 
 which is adopted from the crystal structure of (TMTTF)$_2X$.  
This is an extension of the model studied
  in our previous work~\cite{Yoshioka06JPSJ} 
 to include the lattice dimerization along the chain direction, 
 as well as different Coulomb interactions 
 in the interchain direction. 
As in Ref.~\citen{Yoshioka06JPSJ}, 
 in order to take full account of the 1D fluctuation effects, 
 we treat an effective 1D model by the bosonization theory, 
 by applying mean-field approximation to the interchain interactions. 
Our main results are summarized as follows. 
(i) By deriving an analytical expression of $T_{\mathrm{CO}}$ and computing 
 it by the renormalization group (RG) approach, 
 we find that the lattice dimerization largely suppresses the COI phase 
 and results in the dimer-Mott insulating phase, 
 especially in the critical region. 
(ii) The different interchain Coulomb interactions act as 
 geometrical frustration for the CO state, 
 therefore the COI phase is suppressed 
when these are comparable to each other, 
consistent with previous works~\cite{Seo06JPSJ}. 
(iii) A general formula for the Knight shift 
 below the CO phase transition temperature is derived 
 as a function of the magnitude of order parameter.  
%degree of charge disproportionation. 

We start with the model Hamiltonian
 $H_{\mathrm{Q1D}}= \sum_j H^{j}_{\mathrm{1D}}+H_\perp$, where
 $H^{j}_{\mathrm{1D}}$ represents
the $j$th  extended Hubbard chain
 with finite lattice dimerization $\delta_{\mathrm{d}}$: 
%================== (1) ========================
\begin{align}
H^{j}_{\mathrm{1D}} =&
-t \sum_{i,s}  \left[ 1 + (-1)^i \delta_{\mathrm{d}} \right]
   \left( c^\dagger_{i,j,s} c_{i+1,j,s}^{} + \mathrm{h.c.} \right) 
\nonumber \\ &
 + U \sum_{i} n_{i,j,\uparrow} \, n_{i,j,\downarrow}
+ V \sum_{i} n_{i,j} \, n_{i+1,j},
\end{align}
%==========================================
where $c^\dagger_{i,j,s}$ is the creation operator of an electron 
with spin $s(= \uparrow\!\! / \!\! \downarrow)$
  at the $i$th site on the $j$th chain.
The density operators are
   $n_{i,j,s} \equiv c^\dagger_{i,j,s} c_{i,j,s}^{}- \frac{1}{4}$ and
  $n_{i,j}=n_{i,j,\uparrow}+n_{i,j,\downarrow}$. 
The second term $H_\perp$ expresses the interchain interactions, given by 
%================== (2) ======================== 
\begin{equation}
%HS070409 
H_\perp  = V_\perp \sum n_{i,j} \, n_{i,j'} 
+ V_\perp' \sum n_{i,j} \, n_{i+1,j'},  
\label{eqn:Hperp}
\end{equation}
%========================================
where the $V_\perp$-term denotes
the interchain interaction between electrons
  on the nearest-neighbor sites 
 (along the solid horizontal lines in Fig. \ref{fig1}) and 
the $V_\perp'$-term denotes the interaction 
  for the next-nearest-neighbor sites
(along the dotted diagonal lines).
When $V_\perp=V_\perp'=0$ the model reproduces 
 the above-mentioned 1D dimerized EHM. 

As in Ref. \citen{Yoshioka06JPSJ}, 
 we assume the Wigner crystal-type CO pattern 
  with a 2-fold periodicity along the chain direction~\cite{note1}, 
 and apply the interchain mean-field approach\cite{Scalapino75PRB}
 to $H_\perp$ [eq. (\ref{eqn:Hperp})]. 
The two interchain Coulomb interactions 
 favor different CO along the interchain direction, 
 namely, $V_\perp$ favors the antiphase pattern,  
 $\langle n_{i,j} \rangle  =  (-1)^{i+j} n$, 
 while 
 $V_\perp'$ favors the inphase CO with $\langle n_{i,j}\rangle = (-1)^{i} n$, 
 where $n (>0)$ is the amplitude of CO to be determined self-consistently. 
This results in geometrical frustration arising
  from a competition between two types of CO states~\cite{Seo06JPSJ}.  
The resultant effective 1D Hamiltonian can be written for both cases as 
%=============== (3) ===========================
\begin{align}
H =& 
-t \sum_{i,s} 
 \left[ 1 + (-1)^i \delta_{\mathrm{d}} \right]
\left( c^\dagger_{i,s} c_{i+1,s}^{} + \mathrm{h.c.} \right)
 \nonumber \\ 
&+ U \sum_{i}  n_{i,\uparrow} \, n_{i,\downarrow} 
 + V \sum_{i}  n_{i} \, n_{i+1} 
\nonumber \\
&+ z {\tilde V}_\bot n \sum_i (-1)^i n_i  
+ \frac{1}{2} z N {\tilde V}_\bot n^2,
\label{eqn:eff1D} 
\end{align}
%==========================================
where 
 the chain index $j$ is omitted and $N$ is 
 the total number of sites in a chain. 
The coordination number of adjacent chains, which is
 given by 2 in the system of Fig. \ref{fig1}, 
 is written as $z$.
The effective interchain interaction ${\tilde V}_\perp$ 
 is given by ${\tilde V}_\perp= V_\perp - V_\perp'$, 
 then the interchain frustration simply reduces the 
 effective interchain interaction in the present scheme, 
 therefore acts as destabilizing the CO state, 
 similarly to previous studies on different models~\cite{Seo06JPSJ}.

In order to take fully into account the 1D fluctuation effects,   
we treat the Hamiltonian (\ref{eqn:eff1D}) based on the bosonization theory.
First, we focus on low-energy excitations near the Fermi
wavenumber $\pm k_{\mathrm{F}} = \pm \pi/(4a)$ ($a$: lattice spacing),
by integrating out the high-energy states.\cite{Yoshioka00JPSJ}  
Next, following the standard bosonization theory,
we express the Hamiltonian in the bosonic phase variables 
  and separate it into the charge and spin parts;
  $H=H_\rho+H_\sigma$. 
The spin part $H_\sigma$ takes the same form of  
  the effective theory of the 1D isotropic $S=1/2$ Heisenberg chain, 
  in which the coupling constants depend on $\delta_\mathrm{d}$ as well as $n$.
This indicates that 
  the spin excitation is gapless in the both insulating states, and
the competition between the COI phase and the dimer-Mott state
  is characterized solely by the charge part of the Hamiltonian, $H_\rho$. 
  
Let us first focus on the CO transition temperature $T_{\mathrm{CO}}$. 
Although the phase Hamiltonian is derived  
 for the full order of $n$ by the above operation,
in order to evaluate $T_{\mathrm{CO}}$
the expression up to the first order in $n$
 is sufficient, 
 since a recent numerical study~\cite{Seo07JPSJ} shows that 
 the transition is continuous. 
In this order, the charge part 
  is given by  $H_\rho = \int \mathrm{d}x
 (\mathcal{H}_{\rho 0} + \mathcal{H}_{\rho1})$ where 
%======== (4) (5) ======
\begin{align}
 \mathcal{H}_{\rho 0} = &  \frac{v_\rho}{4 \pi} 
 \left[
\frac{1}{K_{\rho}} (\partial_x \theta_\rho)^2
         + K_{\rho} (\partial_x \phi_\rho)^2
\right] 
\nonumber \\ &  
 - \frac{g_{1/2}}{ (\pi \alpha)^2} \sin 2 \theta_\rho 
+ \frac{g_{1/4}}{2 (\pi \alpha)^2} \cos 4 \theta_\rho,  
\label{eqn:1/4-0} 
\\
\mathcal{H}_{\rho 1} =& 
\frac{z c_2}{\pi a} D
   {\tilde V}_\perp n 
  \cos 2 \theta_\rho 
+  \frac{z}{2a} {\tilde V}_\perp n^2,   
\quad 
\label{eqn:1/4-1}  
\end{align}
%==========
with
$D\equiv (2-\sqrt{1+\delta_{\mathrm{d}}^2})/(1+\delta_{\mathrm{d}}^2)$
 and $c_2 \approx [U/(\sqrt{2}\pi t)] (a/\alpha)^2$.%\cite{Yoshioka06JPSJ}
The quantity $\alpha$ is a short-length cutoff of the order of 
  the lattice constant $a$.
The first term (\ref{eqn:1/4-0}) 
 is the 1D part and 
 takes the same form of the phase Hamiltonian 
 for the 1D dimerized EHM at quarter-filling, 
 derived in Ref. \citen{Tsuchiizu01JPSJ}.\cite{note2}
Here, the effects of finite lattice dimerization, 
 compared with our previous work 
 for $\delta_\mathrm{d}=0$~\cite{Yoshioka06JPSJ}, 
 are 
 (i)  the appearance of the $\sin 2\theta_\rho$ potential and 
 (ii) the modifications of the 
  the coupling constants 
  (e.g., the factor $D$ in the $\cos 2\theta_\rho$ potential).   
The order parameter  $n$ is determined by the 
  self-consistency condition
%======== (6) ======
\begin{equation}
 n = -   \frac{c_2}{\pi} D
\langle \cos 2 \theta_\rho \rangle, 
\label{eqn:n}
\end{equation} 
%===================  
where $\langle \cdots \rangle$ denotes the expectation value 
  with respect to $H_{\rho}$.
Equation (\ref{eqn:n})  is 
 obtained from the stationary condition
 of eq. (\ref{eqn:1/4-1}) and
 takes the same form of eq. (14) in Ref. \citen{Yoshioka06JPSJ}
  except the appearance of the factor $D$.
 The equation to determine  $T_{\mathrm{CO}}$ is given by
%========= (7)
\begin{equation}
 1 = \frac{z c_2^2}{\pi^2 a} D^2 {\tilde V}_\perp
 \int \mathrm{d} x 
\int_0^{1/T_{\mathrm{CO}}} \mathrm{d} \tau
 \, F_{\mathrm{CO}}(x,\tau)|_{T=T_{\mathrm{CO}}},
\label{eqn:TCO}
\end{equation}
%==========     
where
$F_{\mathrm{CO}} (x,\tau) \equiv 
 \langle T_\tau \cos 2 \theta_\rho (x, \tau)
   \cos 2 \theta_\rho (0,0) \rangle_0$
  is the correlation function of the CO order parameter
and $\langle \cdots \rangle_0$ is the expectation value 
with respect to $\mathcal{H}_{\rho 0}$, i.e., 
the Hamiltonian of the purely 1D system.

Here we calculate the correlation function
  based on the RG treatment.
 We also evaluate %another correlation function
 $F_{\mathrm{BO}} (x,\tau) \equiv 
  \langle T_\tau \sin 2 \theta_\rho (x, \tau)
    \sin 2 \theta_\rho (0,0) \rangle_0$,
   which corresponds to the correlation function
  of the bond-order parameter.
Based on the conventional RG treatment, 
  these are given by \cite{Giamarchi1989}
%========== (8)-(9) ===========
\begin{align}
F_{\mathrm{CO}} (\ell)
&= 
\frac{1}{2} \exp 
\left[
- \int_0^{\ell} \mathrm{d} \ell' \left(
4 K_\rho (\ell') + 2 G_{1/4} (\ell')
\right)
\right],  
\label{eqn:cc}
\\
F_{\mathrm{BO}} (\ell)
&= 
\frac{1}{2} \exp 
\left[
- \int_0^{\ell} \mathrm{d} \ell'\left(
4 K_\rho (\ell') - 2 G_{1/4} (\ell')
\right)
\right], 
\label{eqn:ss}
\end{align}
%=============================
where $\ell = \ln (r/\alpha)$ with $r = \sqrt{x^2 + (v_\rho \tau)^2 }$. 
The RG equations of $K_\rho(\ell)$, $G_{1/4}(\ell)$, and $G_{1/2}(\ell)$
    take the same form of eqs. (6)-(8) in Ref. \citen{Tsuchiizu01JPSJ},
 with $K_\rho (0) = K_\rho$, $G_{1/2}(0) = - g_{1/2}/(\pi v_\rho)$, and
$G_{1/4} (0) = g_{1/4}/(2\pi v_\rho)$.
In the 1D case~\cite{Tsuchiizu01JPSJ}, i.e., for $\tilde{V}_\perp=0$,  
 the dimer-Mott insulating state is obtained
  for small intersite Coulomb repulsion ($V<V_{\mathrm{c}}^{\mathrm{1D}}$ 
 when $U$ fixed at a large value),
  where both $G_{1/2}(\ell)$ and $G_{1/4}(\ell)$ become relevant
  under the scaling procedure, 
 while for large intersite repulsion ($V>V_{\mathrm{c}}^{\mathrm{1D}}$), 
   the COI state is obtained in the ground state where
   $G_{1/4}(\ell)$ becomes relevant but 
   $G_{1/2}(\ell)$ irrelevant.
In the present perturbative RG scheme,
  the correct behavior of the correlation functions 
  is not reproduced for large $\ell$
  due to the relevant behavior of
  the nonlinear terms in the low-energy limit. \cite{Giamarchi1989}
In order to avoid this problem,
  we simply stop the scaling procedure at the scale corresponding 
  to the excitation gap $l=l_\Delta$.
For longer distance $l>l_\Delta$, we  assume  that, 
  in the $V>V_{\mathrm{c}}^{\mathrm{1D}}$ case, 
  $F_{\mathrm{CO}} (\ell)$ remains constant while 
  $F_{\mathrm{BO}} (\ell)$ 
   decays exponentially  as a function of $r$
   with correlation length
   $\xi \sim \alpha  \exp (\ell_\Delta)$, 
 and vice versa 
  for  $F_{\mathrm{CO}} (\ell)$ and $F_{\mathrm{BO}} (\ell)$ 
  in the $V<V_{\mathrm{c}}^{\mathrm{1D}}$ case.

%====== Fig.2      
\begin{figure}[t]
\begin{center}\leavevmode
\includegraphics[width=6cm]{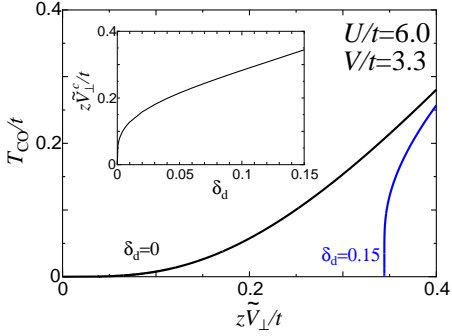}
\end{center} 
\caption{ 
(Color online)
Charge-ordering transition temperature, $T_{\rm CO}$ as a function of $z
 {\tilde V}_\bot /t $ for $\delta_{\rm d} = 0$ and $\delta_{\rm d}= 0.15$
for $V<V_{\mathrm{c}}^{\mathrm{1D}}$.
The inset shows the critical interchain interaction $z {\tilde V}^c_\bot /t $ 
as a function of the dimerization $\delta_{\rm d}$.
}
\label{fig2}
\end{figure}
%===============

By inserting the result of eq. (\ref{eqn:cc})
 into eq. (\ref{eqn:TCO}) with the above procedure, 
 we evaluate $T_{\mathrm{CO}}$. 
In Fig. \ref{fig2},
  we show the transition temperature as a function of 
  $z {\tilde V}_\perp/ t$
   for $\delta_{\mathrm{d}}=0$ and $\delta_{\mathrm{d}}=0.15$ 
   in the $V<V_{\mathrm{c}}^{\mathrm{1D}}$ case.
For $\delta_{\mathrm{d}}=0$, 
 infinitesimal ${\tilde V}_\perp$ transforms the system from TLL to COI state  
 when the TLL parameter satisfy $1/4 < K_\rho < 1/2$, 
 i.e., in the critical region.~\cite{Yoshioka06JPSJ}
The present results show that the dimerization suppresses the transition temperature
   $T_{\mathrm{CO}}$ drastically. 
It is noticeable that a finite amount of 
 $\tilde V_\perp$ is
 necessary for the appearance of the COI state, 
 for which its critical value  
  ${\tilde V}_\perp^{\mathrm{c}}$ is shown in the inset of Fig. \ref{fig2}.    
For  ${\tilde V}_\perp \gtrsim {\tilde V}_\perp^{\mathrm{c}}$, 
the $\tilde V_\perp$ dependence of the 
  transition temperature is expressed as
%---------- (10) -------
\begin{align}
T_{\rm CO} \sim \frac{\Delta_{\rm Mott}}{\ln \left(
 \frac{\tilde V_\bot^c}{\tilde V_\bot-\tilde V_\bot^c}\right) }
, 
\end{align}
%-----------------------
where $\Delta_{\rm Mott}$ is the Mott gap in the absence of the
interchain interaction. 
This formula can be derived by noting that $F_{\mathrm{CO}}$ decays 
  exponentially.

%====== Fig.3
\begin{figure}[t]
\begin{center}\leavevmode
\includegraphics[width=5.7cm]{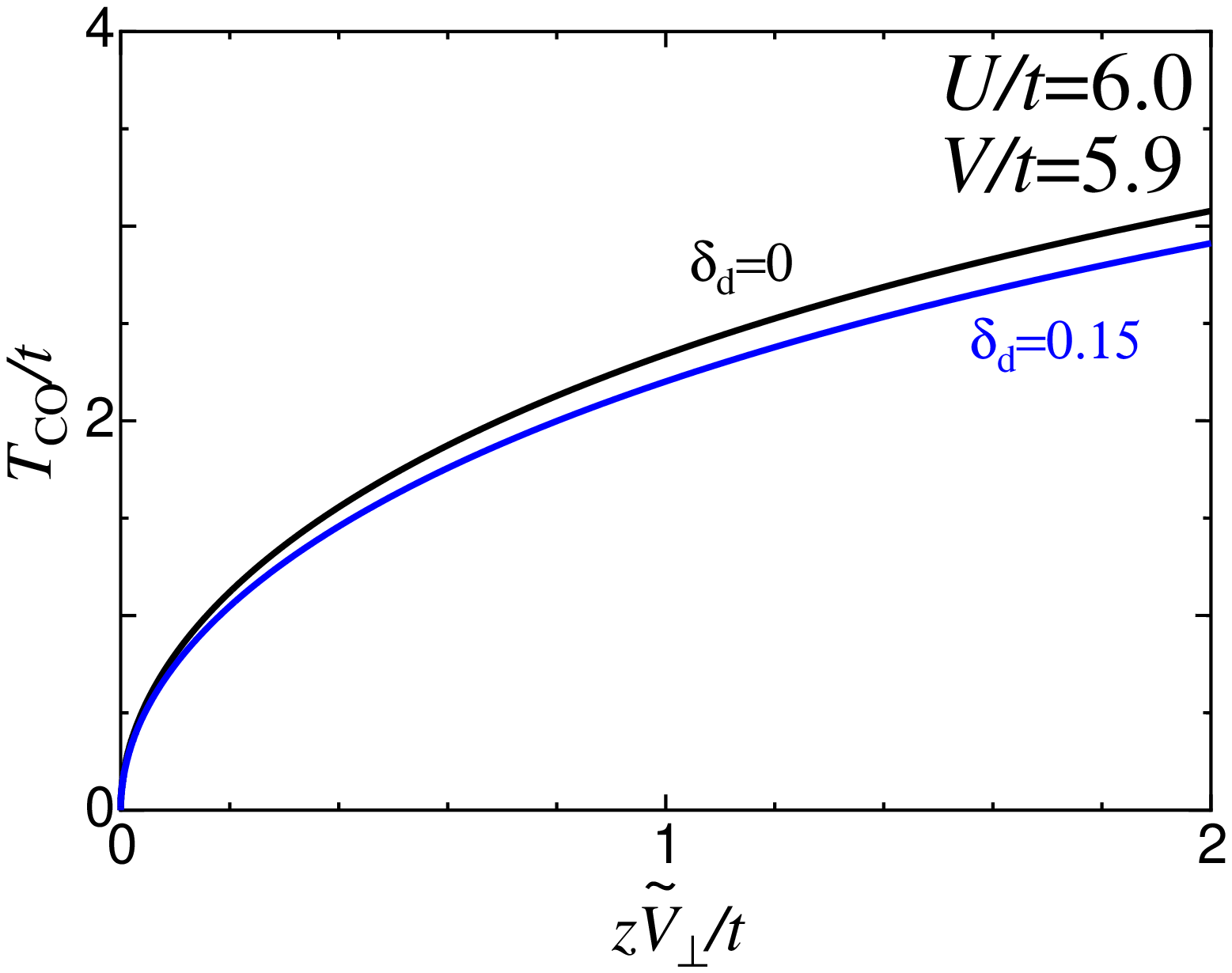}
\end{center} 
\caption{
(Color online)
Charge-ordering transition temperature, $T_{\rm CO}$ as a function of $z
 {\tilde V}_\bot /t $ for 
$\delta_{\mathrm{d}} = 0$ and $\delta_{\mathrm{d}} = 0.15$ 
for $V>V_{\mathrm{c}}^{\mathrm{1D}}$.
}
\label{fig3}
\end{figure}
%===============

In Fig. \ref{fig3}, $T_{\mathrm{CO}}$ 
 for the $V>V_{\mathrm{c}}^{\mathrm{1D}}$ case is shown, 
 where, on the other hand, the COI state is stabilized  even for $\tilde V_\perp=0$. 
Again, the dimerization suppresses $T_{\rm CO}$. 
 However, for a fixed value of $\delta_{\rm d}$, 
 the critical value ${\tilde V}_\perp^{\mathrm{c}}$ 
 of the effective interchain interaction $\tilde V_\perp$ 
 for the appearance of the CO state 
 is now always zero, i.e., 
 infinitesimal interchain interaction leads to
 finite $T_{\mathrm{CO}}$.
The $\tilde V_\perp$ dependence of the transition temperature is 
 now given by
$T_{\mathrm{CO}} \propto \tilde V_\bot^{1/2}$
as long as $T_{\mathrm{CO}} \ll \Delta_{\mathrm{CO}}$ 
with $\Delta_{\mathrm{CO}}$ being the charge gap of the purely 1D COI.

\begin{figure}[b]
\begin{center}\leavevmode
\includegraphics[width=6.4cm]{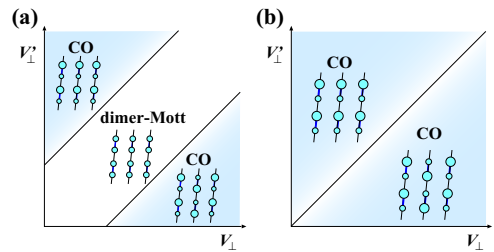}
\end{center}
\caption{
(Color online)
Schematic phase diagrams of model (\ref{eqn:eff1D})
 on the plane of 
 the interchain interactions $V_\perp$ and $V_\perp'$,
  for the $V<V_{\mathrm{c}}^{\mathrm{1D}}$ case (a) and for
  the $V>V_{\mathrm{c}}^{\mathrm{1D}}$ case (b). 
CO and dimer-Mott stand for charge-ordered insulating phase with $T_{\mathrm{CO}}>0$
 and dimer-Mott insulating phase, 
 respectively. 
}
\label{fig:phase}
\end{figure}

Summarizing the above results, 
schematic phase diagrams on the $V_\perp$-$V_\perp'$ plane 
  are shown in Fig. \ref{fig:phase} where  
the shaded CO regions denote $T_{\mathrm{CO}} > 0$.  
In the  $V< V_{\mathrm{c}}^{\mathrm{1D}}$ case (a),
between the
  anti-phase and in-phase CO states, there appears  
  the dimer-Mott insulating phase where CO is absent.  
On the other hand, in the $V> V_{\mathrm{c}}^{\mathrm{1D}}$ case
(b),  there is a direct transition between the two CO states;
at the boundary, $T_{\mathrm{CO}} = 0$, 
i.e., CO is realized only at the ground state.    

%===============
%===============

Next, we focus on the critically fluctuating line, 
 $V=V_{\mathrm{c}}^{\mathrm{1D}}$.  
The quantum phase transition in the 1D case (i.e., $\tilde V_\perp=0$)
 is known to be the Ising-type, as mentioned. 
On this line 
 the finite-temperature CO phase transition
% due to finite interchain coupling
for ${\tilde V}_\perp \neq 0$ 
shows nontrivial behavior, since
 the correlation function of the CO order parameter $F_{\mathrm{CO}} (x,\tau)$ exhibits 
  a power-law of $r^{-1/4}$. 
At finite temperature, the analytical form can be obtained 
as\cite{Fabrizio00NP}
%---------- (11) -------
\begin{align}
&F_{\mathrm{CO}} (x,\tau) = \langle T_\tau \cos 2 \theta_\rho (x, \tau)
   \cos 2 \theta_\rho (0, 0) \rangle
 \nonumber \\
& \quad\propto 
\left[
\frac{(\pi T)^2} {\sinh \left[\pi T (x - \mathrm{i} \tau)
 \right] \sinh \left[ \pi T (x + \mathrm{i} \tau) \right]}
\right]^{1/8}.
\end{align}
%-----------------------
By inserting this into eq. (\ref{eqn:TCO}),
we find that the $\tilde V_\perp$
  dependence of the transition temperature is given by
  $T_{\rm CO} \propto \tilde V_\perp^{4/7}$.

Now we turn to the COI phase below the transition temperature
 $T_{\mathrm{CO}}$ and derive a useful formula for the 
 relation between splitting in the NMR spectrum~\cite{Hiraki1998PRL,Chow2000PRL,TakahashiReview} 
 and the CO amplitude $n$. 
In the presence of the Wigner crystal-type CO, 
 an effective alternating site potential $zV_\perp n (-1)^i n_i$ 
 (its wavenumber is $q=\pi/a = 4k_{\mathrm{F}}$) 
 is applied to the system [see, eq.(3)]
 and results in folding of the one-particle dispersion relation. 
Then the $4k_{\mathrm{F}}$-component
 of the density operators is expressed in part by 
 the $q=0$ density operators around the Fermi energy, namely, 
the slowly-varying component of 
the charge/spin density becomes a part of its $4k_{\mathrm{F}}$ component.
This is directly related to the splitting in the NMR spectrum  
 since the response under uniform magnetic field at each site becomes disproportionated 
 with a period of two sites~\cite{Yoshioka06JPSJnmr} 
 reflecting the Knight shift.  
In the phase representation,
 the staggered component of the spin density
 $m_{4k_{\mathrm{F}}} (x) $ is proportional to 
 the slowly varying component of the spin density
 $m_0 (x) = \partial_x \theta_\sigma(x)/\pi$ as 
%============= (12) ============
\begin{align}
 m_{4k_\mathrm{F}} (x) = - (-1)^i
 \frac{z \tilde V_\perp n}
{\sqrt{2t^2 (1+\delta_{\mathrm{d}}^2) + (z \tilde V_\perp n)^2}}
 \frac{\partial_x \theta_\sigma(x)}{\pi}.   
\end{align}
%===============
From this relation, 
 the Knight shift at the $i$th site can be expressed as
%============= (13) ============
\begin{align}
 S_i (T) \propto \chi_\sigma (T) 
\left[
1 - (-1)^i \frac{z \tilde V_\perp n}
{\sqrt{2t^2 (1+\delta_{\mathrm{d}}^2) + (z \tilde V_\perp n)^2}}  
\right],  
\label{eq:S_stag}
\end{align}
%===========================
where $\chi_\sigma (T)$ is the magnetic susceptibility. 
This verifies that 
  the average of the Knight shift is proportional to the magnetic
  susceptibility as observed experimentally.\cite{Hiraki1998PRL} 
Furthermore we find that 
 the ``relative shift'' $(S_1 - S_0)/(S_1 + S_0)$
  can be expressed as
%=============== (14) ========
\begin{equation}
 \frac{S_1 - S_0}{S_1 + S_0} = \frac{z \tilde V_\perp n}
{\sqrt{2t^2 (1+\delta_{\mathrm{d}}^2) + (z \tilde V_\perp n)^2}}.
\label{eqn:estimation}   
\end{equation}
%=======================
We note that 
the \textit{intrachain} interactions 
do not explicitly appear in this expression,
  but  they are embedded 
  in the magnitude of the order parameter $n$. 
Experimentally, the order parameter of the CO phase 
 has been sometimes estimated so far by simply assuming
 that the Knight shift is proportional to an amount of the charge located
 on the site, i.e., 
 $ (S_1 - S_0)/(S_1 + S_0) \to 2n $.~\cite{Hiraki1998PRL} 
However, 
the value of $n$ obtained in such a way can have been misestimated. 
% the correct formula is given by eq. (\ref{eqn:estimation}) and 
% the experimentally-estimated value of $n$, 
% when the splitting in the Knight shift is referred to, 
% can have been misestimated.

We compare the present results with recent 
experimental observations 
by dielectric permittivity\cite{Nad05JP} 
and ESR\cite{Furukawa05JPSJ}
that 
$T_{\mathrm{CO}}$ for 
some (TMTTF-$d_{12}$)$_2X$ salts 
composed of the perdeuterio-TMTTF 
are increased.  
The X-ray analysis for the TMTTF-$d_{12}$ and TMTTF-$h_{12}$ compounds
has revealed that 
the deuteration leads to suppression of the lattice 
  dimerization, \cite{Furukawa05JPSJ} 
 which is consistent with the present results.
The rather large rise in $T_{\mathrm{CO}}$ of about 10 percent 
compared to the small change in dimerization suggests that 
the systems are in the critical region, $V < V_{\rm c}^{\rm 1D}$.

The anomalous behavior of the dielectric
 constant, proportional to $|T-T_{\mathrm{CO}}|^{-1}$, 
 observed in the TMTTF compounds\cite{Monceau2001PRL,NadReview}, 
 has been theoretically discussed 
 based on the dimerized EHM 
 in the purely 1D case\cite{Monceau2001PRL,Tsuchiizu02JPCS}.   
It is suggested that at $T=0$ the dielectric constant
   shows a divergent behavior 
   at the boundary between the Mott insulator and the COI, i.e., 
   $V=V_{\mathrm{c}}^{\mathrm{1D}}$. 
If the interchain coupling is taken into account 
as in our work here, 
this quantum phase transition would be changed into a finite-temperature
 transition and we expect that 
 the dielectric constant should show a similar anomalous behavior 
  at $T_{\mathrm{CO}}$. 
The investigation of such a possibility
  in the present scheme remains for a future study.

In conclusion, we have discussed the 
  competition between the charge-ordered insulating state
   and the dimer-Mott insulator at  
  finite temperatures
  in a quasi-one-dimensional extended Hubbard model at quarter filling.
Frustration in the interchain Coulomb repulsion reduces
  the effective interchain interaction and therefore decreases 
 the charge-ordering transition temperature $T_{\mathrm{CO}}$. 
The lattice dimerization also reduces $T_{\mathrm{CO}}$, 
  while a finite strength of the effective interchain interaction is necessary
  to transform the system from the dimer-Mott insulating state 
  into the charge-ordered insulating state.
Based on the present analysis, we have derived 
 a useful formula for estimating   
 the magnitude of the order parameter of the CO phase experimentally.

\section*{Acknowledgment}
The authors thank H. Fukuyama and Y. Suzumura for valuable comments, 
and Y. Otsuka for stimulating discussions. 
This work was supported by a Grant-in-Aid for Scientific Research on
Priority Area of Molecular Conductors (Nos.18028018, 18028026, and 18740221)
  from the
Ministry of Education, Culture, Sports, Science and Technology.


\begin{thebibliography}{99}
\bibitem{noteCODM}
In the presence of alternation in transfer integrals, 
the CO state may be called a co-existence of the dimer-Mott insulator and CO 
since dimerization is always present; 
in this paper we call this simply as the CO state.
\bibitem{note1}
We do not consider the $2k_{\mathrm{F}}$-CDW state with a 4-fold periodicity along the chains
since we are interested in stability of Wigner crystal-type charge ordering.
\bibitem{note2}
The notation of the coupling constant $g_{1/2}$ differs from that in 
 Ref. \citen{Tsuchiizu01JPSJ}.
The coefficient of $\sin 2 \theta_\rho$ in eq. (\ref{eqn:1/4-0}),
  should be multiplied by $-\frac{1}{2}$ in order to reproduce 
   the Hamiltonian in Ref. \citen{Tsuchiizu01JPSJ}. 
The change in the sign is due to the choice of the Klein factor.
\end{thebibliography}
\end{document}